\documentclass[prl,twocolumn,showpacs,preprintnumbers,amsmath,amssymb]{revtex4}% 

\usepackage{graphicx}% Include figure files
\usepackage{dcolumn}% Align table columns on decimal point
\usepackage{bm}% bold math

\begin{document}

%\preprint{APS/123-QED}

\title{Nanosecond microscopy with a high spectroscopic resolution}% Force line breaks with \\

\author{Christoph Heinrich}
\author{Stefan Bernet}%
\author{Monika Ritsch-Marte}%
\affiliation{Division of Biomedical Physics, Innsbruck Medical University, M{\"u}llerstr. 44, A-6020 Innsbruck, Austria}%

%\affiliation{
%This line break forced% with \\}%

\date{\today}% It is always \today, today,

\begin{abstract}
We demonstrate coherent anti-Stokes Raman scattering (CARS) microscopy in a wide-field setup with nanosecond laser pulse excitation. In contrast to confocal setups, the image of a sample can be recorded with a single pair of excitation pulses. For this purpose the excitation geometry is specially designed in order to satisfy the phase matching condition over the whole sample area. The spectral, temporal and spatial sensitivity of the method is demonstrated by imaging test samples, i.e. oil vesicles in sunflower seeds, on a nanosecond timescale. The method provides snapshot imaging in 3 nanoseconds with a spectral resolution of 25 cm$^{-1}$.
\end{abstract}

\maketitle

%\section{\label{sec:level1}First-level heading:\protect\\ The line
%break was forced \lowercase{via} \textbackslash\textbackslash}

%\section{\label{sec:level1}}
%\subsection{\label{sec:level2}Second-level heading: Formatting}
%\subsubsection{\label{sec:level3}Third-level heading: References and Footnotes}

Coherent anti-Stokes Raman scattering (CARS) has established as a spectrally selective method in microscopy, e.g. for imaging structures within biological cells \cite{Duncan,Zumbusch,Xie4}. In contrast to fluorescence microscopy, it is not necessary to stain the sample with fluorescent dyes. Instead, excitation is performed with a pair of laser pulses which have a frequency difference corresponding to a Raman transition of the selected structures. This results in coherent emission of light at the so-called anti-Stokes frequency. The blue-shifted signal can be easily separated from the excitation light by corresponding high-pass filters. Typically, a confocal microscopy setup is used, where the two excitation pulses are collinearily focused  at a single spot of the sample, where the emitted anti-Stokes signal is detected from. Imaging is then performed by sampling the specimen point-by-point and reconstructing its shape by the computer. For this purpose, the pulses are provided by pico- or femtosecond laser systems which provide a high pulse intensity required for the nonlinear process at a sufficiently high repetition rate for fast sampling. Advantages of such a confocal setup consist in its high spatial resolution, which is similar to confocal fluorescence microscopy, in the possibility of three-dimensional sectioning of the object volume, and in the chemical selectivity, allowing to image specific substances by controlling the frequency difference of the excitation pulses \cite{Xie5}.

However, here we demonstrate the advantages of a different CARS microscopy setup in a wide-field imaging geometry \cite{Heinrich} using nanosecond excitation pulses. Such a system is complementary to typical confocal scanning setups: On the one hand, our wide-field system offers a lower spatial resolution, because it lacks the spatial filtering capabilities of confocal microscopy. On the other hand, it promises faster image acquisition, since the whole sample is recorded without scanning, and it can principally provide a better chemical resolution due to the smaller bandwidth of the nanosecond excitation pulses as compared to pico- or femtosecond systems. 

Here we demonstrate ultra-fast microscopic imaging of cell components with a single pair of excitation pulses on a nanosecond timescale, with a spectral resolution on the order of a few wavenumbers, corresponding to a wavelength bandwidth in the sub-nanometer regime  - which provides a three-dimensional sectioning capability (vertical resolution) on the order of one micron. As a test sample we use cells of sunflower seeds, where included oil-filled vesicles are imaged using the Raman-transition of the contained linoleic acid molecules at the 2870 cm$^{-1}$ aliphatic CH-stretch-vibration to generate a CARS signal.

For this purpose we use an excitation geometry already presented in a previous publication \cite{Heinrich}, which is similar to a combination of epi-fluorescence, and ultra-dark-field microscopy, and which satisfies the required "wave-matching condition" over the whole sample area simultaneously. This setup has the additional advantage that it collects all signal light, even if the microscope objective has only a small numerical aperture, and that it is inherently background-free, i.e. the optical beam path guides only the CARS-signal photons directly to the CCD camera.

\begin{figure}
\includegraphics[width=\linewidth]{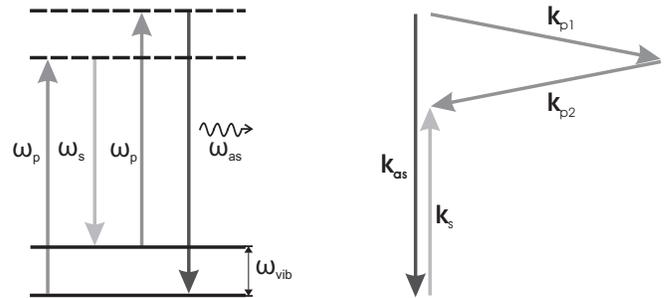}% Here is how to import EPS art
\caption{\label{fig:phasematch}  Left: CARS transition scheme. Two incident pump photons with a frequency of $\omega_{p}$ are converted into an outgoing Stokes- ($\omega_{s}$) and an anti-Stokes ($\omega_{as}$) photon in the vicinity of a Raman active molecule with a vibrational transition frequency of $\omega_{vib}$. The process is stimulated by an external Stokes beam.  Right: Extremely folded CARS wave-matching geometry \cite{Mueller,Heinrich} with two pump photons (wave vectors $\vec{k}_{p1}$ and $\vec{k}_{p2}$) being converted into a Stokes ($\vec{k}_{s}$) and an anti-Stokes ($\vec{k}_{as}$) photon which are counter-propagating.}
\end{figure}

In general, the generation of a CARS image in a wide-field setup is a non-trivial task, since two conditions have to be satisfied in the extended interaction region \cite{Eesley}. First, the CARS transition scheme shown on the left hand side of Fig.~\ref{fig:phasematch} sketches the requirement of energy conservation: In the vicinity of a Raman-active molecule with a vibrational transition at a frequency $\omega_{vib}$, two incoming pump photons with frequency $\omega_{p}$ are converted into an outgoing Stokes-, and an anti-Stokes photon (which is the desired signal), such that the corresponding frequencies are determined by: $\omega_{s}=\omega_{p}-\omega_{vib}$ and $\omega_{as}=\omega_{p}+\omega_{vib}$, respectively. The efficiency of this process is strongly increased by providing an external Stokes beam for stimulated emission of the Stokes photon.

The second condition is based on the required momentum conservation between the 4 interacting photons. It requires that the sum of the wave-vectors of the two pump photons equals the sum of the Stokes and the anti-Stokes wave-vectors, i.e. $\overrightarrow{k_{p1}} + \overrightarrow{k_{p2}} = \overrightarrow{k_{s}}+\overrightarrow{k_{as}}$. 
The two conditions of energy and momentum conservation have a trivial solution in a non-dispersive medium: There the 4 interacting photons can travel collinearily. In a liquid or a solid, however, the medium is dispersive and the wave-matching condition can only be satisfied at special beam propagation directions. One solution - which we use in our setup - is indicated in Fig.~\ref{fig:phasematch} on the right hand side: Our goal was to find a geometry where the anti-Stokes signal beam counter-propagates with respect to the incident Stokes beam. This can be achieved if the two corresponding pump photons come from two almost opposite directions which are nearly perpendicular to the directions of the Stokes- and anti-Stokes beam. It turns out that the angle $\alpha$ between the direction of the anti-Stokes beam $\overrightarrow{k_{as}}$ and each of the two pump-beams $\overrightarrow{k_{p1}}$ and $\overrightarrow{k_{p2}}$ has to satisfy \cite{Heinrich}:

\begin{equation}
\cos(\alpha)= \frac{(k_{as}-k_{s})}{2 k_p} \approx \frac{\omega_{vib}}{\omega_p}
\label{eq:alpha}\; ,
\end{equation}
The approximation holds for sufficiently small dispersion.

\begin{figure}
\includegraphics[width=\linewidth]{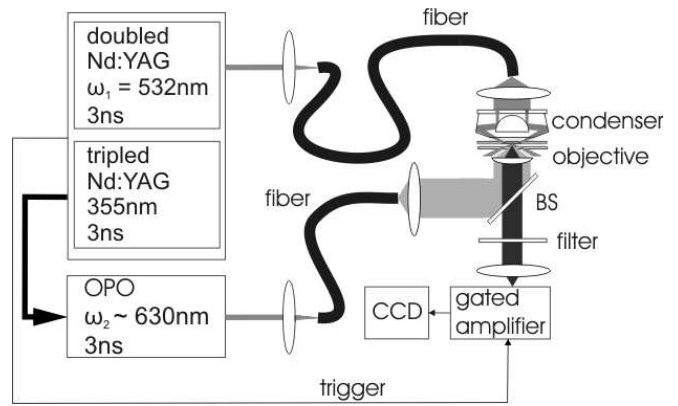}% Here is how to import EPS art
\caption{\label{fig:setup} Experimental implementation. The pump pulse enters the sample from above under a flat angle on the surface of a cone through an ultra-dark-field condenser with a high numerical aperture. The Stokes pulse comes from below through the microscope objective, and the emitted signal pulse counter-propagates back through the same microscope objective and a dichroic beam splitter to an intensified, gated CCD imaging system.}
\end{figure}

Fig.~\ref{fig:setup} is a sketch of our experimental setup: A sample is sandwiched between two cover-slips on a microscope stage (Zeiss Axiovert 135). The 532 nm pump-pulse is emitted from a doubled ND:YAG laser system (with a pulse width of 3 ns), and guided by a multi-mode-fiber (diameter 1000 $\mu$m) to the input of an oil-immersion ultra-dark-field condenser with a numerical aperture range between 1.2 and 1.4. The homogeneously illuminated fiber output is sharply imaged in the sample plane as a circular area with a diameter of approximately 60 microns. The directions of the pump photons lie on the surface of a cone with an opening half-angle in the range of $64^{o}-90^{o}$ degrees (in water between the two coverslips). This available angular range allows to satisfy the wave-matching condition for CARS transitions in a wave-number range between 0 and 8000 cm$^{-1}$.

The Stokes-pulse is emitted from an optical parametric oscillator (OPO) which is pumped by the tripled ND:YAG pulse of the pump laser. The wavelength of the Stokes beam can be scanned continuously in a range between 420 nm and 2600 nm. The bandwidth of the OPO depends on its emission wavelength, and in our case (around a wavelength of 628 nm)it is  on the order of 30 cm$^{-1}$. The pulse is guided by another multi-mode fiber (diameter 600 microns) to the epi-fluorescence input of the microscope. The homogeneously illuminated fiber output is also imaged through a dichroic mirror and an oil-immersion microscope objective (magnification 40 $\times$ with a numerical aperture of 0.65) into the sample plane in a circular area with a diameter of approximately 90 microns, where it arrives at the same time and overlaps with the image of the pump-beam fiber. The Stokes pulse frequency can be scanned to match a Raman transition of structures within the sample. In this case, an anti-Stokes signal is emitted, and - due to the spatial coherence of the generated CARS signal - it travels directly to the microscope objective, from where it is collected and transmitted by the dichroic mirror to an imaging system consisting of an intensified CCD camera. The microchannel intensifier is gated with a trigger pulse of the pump laser system, such that it records only signals during the duration of the excitation pulses, thus suppressing background and luminenscence light. The intensity of the unpolarized pump- and Stokes pulses in the sample plane are on the order of one micro-Joule at a repetition rate of 10 Hz. We stay below the damage threshold of the optics by using multi-mode fibers which cause a phase-front distortion preventing the formation of a diffraction-limited focus somewhere in the beam path or at the sample.
This setup has the advantage that it collects all signal light. Thus, microscope objective with a small numerical aperture can be used. Furthermore, the method is inherently background-free, i.e. the optical beam path guides only the CARS photons directly to the CCD camera, whereas the pump-pulse misses the microscope objective due to its flat angle caused by the ultra-dark-field condenser.

\begin{figure}
\includegraphics[width=\linewidth]{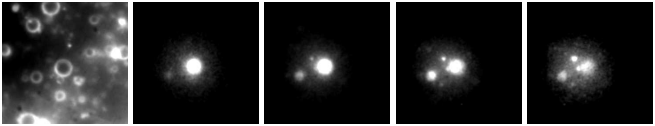}% Here is how to import EPS art
\caption{\label{fig:zaxis} Images of a test sample consisting of a cut through a sunflower seed immersed in water. Image frame diameter is 130 $\mu$m. Left: Dark-field image with white light illumination. Next four pictures: Resonant CARS images captured in 60 s at different vertical positions of the dark-field condenser (displaced by 2 microns from image to image), demonstrating the vertical sectioning capability of the setup.}
\end{figure}

A sample image recorded with this setup is displayed in Fig.~\ref{fig:zaxis}. The left image shows a dark-field exposure (taken with white light) of a cut through a sunflower seed immersed in water and sandwiched between two glass coverslips. The circular areas consist of oil-filled vesicles containing Raman-active linoleic acid. The following sequence of 4 exposures shows CARS images taken at a Stokes wavelength (627.9 nm) matched to the CH-stretch vibration Raman resonance of linoleic acid at a wave-number of 2870 cm$^{-1}$. The images were taken at different heights of the ultra-dark-field condenser above the sample, with a vertical spacing from exposure to exposure of 2 microns. Each image was taken by integrating over a number of 600 excitation pulses, corresponding to a 60 s exposure at our pulse repetition frequency of 10 Hz. The images demonstrate that the oil-filled vesicles are excited to emit a CARS signal (at a wavelength of 461.5 nm) in a region with a diameter of approximately 50 microns, where the two excitation beams overlap. The sequence of images taken at different vertical positions of the dark-field condenser reveals a remarkably high vertical resolution of the system, since images taken at a distance of 2 microns already appear completely different. This vertical resolution is based on the high numerical aperture of the ultra-dark-field condenser, and it can be used for sectioning of a specimen in order to produce a three-dimensional sample profile.

\begin{figure}
\includegraphics[width=\linewidth]{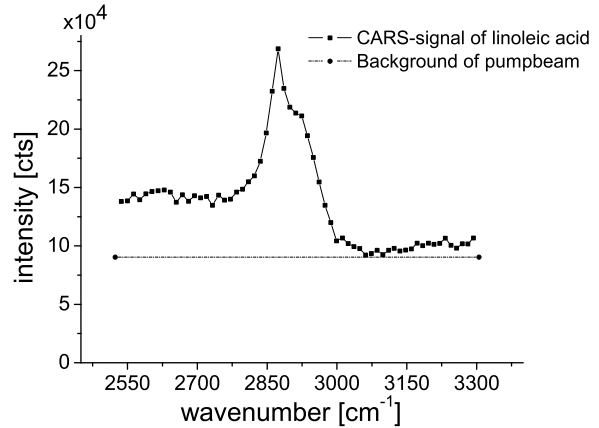}% Here is how to import EPS art
\caption{\label{fig:spectrum}  Spectroscopic CARS signal of the aliphatic CH-vibration of linoleum acid, detected in a region around the resonant CARS wavelength of 461.5 nm after excitation with 532 nm (fixed), and a tuneable wavelength (OPO) around 628 nm. The average signal intensity in the center of images (like those plotted in Fig.~\ref{fig:zaxis}) is recorded at various Stokes-beam wavelengths and plotted as a function of the frequency difference between the two excitation pulses converted to wave numbers. The resulting curve exhibits the typical spectral asymmetry of spectroscopic CARS data, and demonstrates a frequency resolution on the order of 25 cm$^{-1}$.}
\end{figure}

In order to verify the CARS-origin of the detected signal, we measured a spectrum of the signal by recording a number of images at different wavelength of the Stokes-pulse. The camera was used as an intensity detector by integrating the CCD counts in the CARS-active area within the images. The results are drawn in Fig.~\ref{fig:spectrum}.

The data demonstrate the expected lineshape of the CARS resonance at 2870 cm$^{-1}$. It shows the typical asymmetry based on the interference of the resonant CARS signal with a non-resonant (but coherent) signal from the background \cite{Eesley}. Furthermore, the spectrum demonstrates the wavelength resolution of the setup which is on the order of 25 cm$^{-1}$, corresponding to a wavelength bandwidth of less than 1 nm, which is better resolved than typical CARS images taken with ultra-short laser pulses. The spectroscopic resolution is limited by the bandwidth of our currently used broad-band OPO which is 20-30 cm$^{-1}$ in the respective wavelength range, whereas the pump beam has a spectral bandwidth of only 0.03 cm$^{-1}$. Although a spectral resolution below 10 cm$^{-1}$ has been already obtained with optimized picosecond laser systems \cite{Xie4,Cheng1,Mueller1,Potma1}, the use of a narrow-band nanosecond-OPO in our setup would promise an easily achievable spectral resolution on the order of 0.1 cm$^{-1}$.

\begin{figure}
\includegraphics[width=\linewidth]{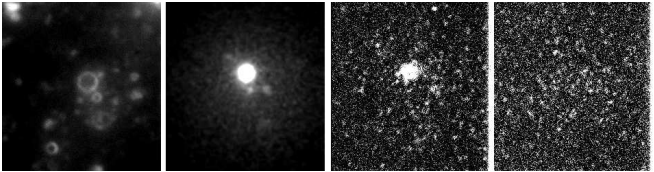}% Here is how to import EPS art
\caption{\label{fig:linol} Time resolution of wide-field CARS imaging: The left picture shows a sunflower-seed sample under dark-field illumination with white light, showing an oil-drop in the middle. The next picture shows a resonant CARS image of the sample, taken with an exposure time of 10 s by integrating over 100 excitation pulses. The third picture shows a single exposure resonant CARS image, recorded in 3 nanoseconds. The final picture again shows a single exposure CARS image of the sample, however taken under non-resonant conditions after detuning the Stokes-laser. Clearly, the spectroscopic resolution of the system is still present even under single shot excitation.}
\end{figure}

In order to check the temporal resolution of our setup we performed imaging experiments with a decreasing number of pulses, down to single-shot experiments. The results of such an experiment are shown in Fig.~\ref{fig:linol}. The left image again shows a dark-field exposure taken with white light illumination. It shows a drop of sunflower-seed oil in the lower right image section. The following image shows a CARS-exposure of the same sample, integrating over 100 shots of the excitation pulses with the Stokes-laser again adjusted for resonance at the 2870 cm$^{-1}$ CH-vibration. The CARS process leads to a high contrast image of the oil drop.  The next picture of the sequence shows again a resonant CARS image of the drop, however, now taken with only one \emph{single} pair of excitation pulses. The contrast is sufficiently high for the CARS-active drop to be recognized and distinguished from its surroundings. Although the low signal intensity leads to a reduced spatial resolution, the position and the approximate size of the drop are clearly detected. As a test, the next image was taken under the same conditions, however with the Stokes-laser detuned from the CARS resonance by 7 nm. As expected, the signal disappears. The comparison of the resonant and non-resonant experiments thus demonstrates that the spectroscopic CARS resolution is already present at single shot exposures. For future experiments, we expect much improvement by using optimized optical components for a larger excitation power and a higher signal transmission.

Our experiment is a first demonstration of an improvement in image exposure time of nine orders of magnitude compared to the fastest other CARS setups, which use confocal scanning methods. Possible applications of such a snapshot CARS imaging could consist in the study of fast processes in biological systems, like electro-physiological processes (nerve signaling, cilia mechanisms), in imaging the progress of diffusion processes \cite{Wiersma}, or to study chemical reactions in micro-fluids. On the other hand, the high spectroscopic resolution of nanosecond CARS imaging will make it possible to map the spatial distribution of different chemical species within biological samples in the highly selective fingerprint regime of organic molecules (mainly in the vicinity of 1200 - 1600 cm$^{-1}$) \cite{Hashimoto}. For example, imaging a sample at a sequence of selected CARS transition frequencies of different organic molecules can be followed by numerical image processing to produce a pseudocolour map of the chemical composition of the sample. The very high spectral resolution which can be principally obtained with nanosecond excitation has its major applications in material research, where sub-wavenumber vibrational features are reported for example in crystalline structures of semiconductors \cite{Theis}, deposited diamond coatings \cite{Stuart}, or crystalline fullerenes \cite{Horoyski}. Altogether, we feel that the combination of the two features of our setup, i.e. ultra-fast imaging and a high spectral resolution, is an ideal complement to the superior spatial resolution offered by confocal scanning CARS systems.

%\appendix
%\section{Appendixes}

%\newpage %Just because of unusual number of tables stacked at end
%\bibliography{cars}% Produces the bibliography via BibTeX.

\end{document}